\documentclass[aps,prl,onecolumn,preprintnumbers,groupedaddress,showpacs]{revtex4}
\usepackage{graphicx}
\usepackage{dcolumn}
\usepackage{bm}

\begin{document}
\def\coup{K}
\def\heli{\Upsilon}
\def\be{\begin{eqnarray}}
\def\ee{\end{eqnarray}}
\preprint{LM8537}

\title{New universality class for the three-dimensional $XY$
model with correlated impurities:
application to $^4\!$He in aerogels}

\author{C. V\'asquez R.$^{1,2}$, R. Paredes V.$^1$, A. Hasmy$^1$, and R.
Jullien$^2$}

\affiliation{$^1$LFESD, Centro de F\'{\i}sica, IVIC,
Apartado 21827, Caracas 1020A, Venezuela}
\affiliation{$^2$LDV, Universit\'e Montpellier II, CC069
Place Eug\`ene Bataillon, 34095 Montpellier, France}

\begin{abstract}
Encouraged by experiments on $^4\!$He in aerogels,
we confine planar spins in the pores of simulated aerogels (DLCA) in order
to study the effect of quenched disorder on the critical behavior
of the three dimensional $XY$ model.
Monte Carlo simulations and finite size scaling are used to determine
critical couplings $K_c$ and exponents.
In agreement with experiments, clear evidence of change in the thermal critical
exponents $\nu$ and $\alpha$ is found at non-zero volume fractions
of impurities. These changes are explained in terms of {\it hidden}
long-range correlations within disorder distributions.
\end{abstract}
\date{\today}
\maketitle

The superfluid ($\lambda$) transition of $^4\!$He in porous media has been
subject of exciting work due to the change of critical exponents observed
in many experiments \cite{yoon,chan,larson}.
Close to the transition temperature ($T_\lambda$),
the relative superfluid density scales as
$\rho_s/\rho\!\sim\! |T\!-\!T_\lambda|^\zeta$
where $\zeta \!\simeq\! 0.671$ is the critical exponent for the pure system
\cite{lipa}.
The correlation length behaves as $\xi\!\sim\! |T-T_\lambda|^{-\nu}$
where the exponent $\nu$ is given by the hyperscaling relation
\cite{josephson,mefisher}
\be
\label{zeta}
\zeta=(d-2)\nu\,,
\ee
with $d$ being the dimension of the system.
Thus, $\zeta\!=\!\nu$ for bulk $^4\!$He at the $\lambda$-transition.
Experiments on  $^4\!$He embeded in porous media such as silica aerogels
and xerogels report an increase of $\zeta$ up to
$0.81$ and $0.89$, respectively \cite{yoon,chan}.
Similar studies with other kind of materials, e.g., vycor and porous gold
\cite{chan,crowell}, do not reveal any change
of criticality. It has been argued \cite{chan}
that discrepant results between the two families of materials obey
different regimes in spatial correlations.

Harris criterion \cite{harris} establishes that short-range correlated
(SRC) impurities do not affect criticality
in systems with a negative specific heat exponent $\alpha$.
The extended criterion due to Weinrib and Halperin \cite{weinrib} (WH)
states that disorder can be relevant
when it exhibits long-range correlations (LRC), 
even in the $\alpha\!<\!0$ case.
Quenched defects obeying power-law correlations, $g(r)\!\sim\! r^{-a}$ at long
distances $r$, are relevant
under the condition $a\nu-2<0$ for $a\!<\!d$ (LRC condition)
where $\nu$ is the exponent for the original pure system.
Prudnikov {\it et al.} \cite{prudnikov}
obtained more precise results for thermal ($\nu$) and magnetic ($\eta$)
exponents of general $N$-vector models in $d\!=\!3$.
One of their main results is that WH's heuristic prediction,
i.e., that $\nu_{LRC}\!=\!2/a$ would be the exponent
for the impure system,
does not hold in their more accurate field-theoretical description.
In practice, the $\lambda$-transition of $^4\!$He
exhibits $\alpha\!\simeq\! -0.013$
\cite{lipa}, so it has been considered
as an ideal system to test these criteria \cite{chan}.

Moon and Girvin \cite{moon} (MG) have notably confirmed
WH predictions, performing Monte Carlo
simulations of the three-dimensional $XY$ model (3DXY), which is
in the same universality class as the superfluid bulk $^4\!$He.
These authors studied the 3DXY model in the presence of correlated disorder,
the fractal incipient percolating cluster,
and found $\zeta \simeq 0.722$ in contrast to the pure 3DXY value.
They also studied the 3DXY model with random bond dilution
(which is SRC), and observed no changes in criticality, 
consistent with Harris criterion.

Porous materials used in $^4\!$He experiments where
changes of criticality occur are obtained through sol-gel processes.
In principle, they exhibit fractal features only at finite scales
($10\mbox{nm}\! <\! r\!<\!\Lambda \!\simeq\! 100\mbox{nm}$)
and low concentrations (porosities $\varphi\!\geq\! 90\%$), appearing
homogeneous at large scales or large concentrations \cite{hasmy}.
According to theory, no change of universality
should be expected for the superfluid transition in the latter cases:
after the WH criterion, as $T\!\to\! T_c$ ($\xi\!\to\!\infty$),
disorder should not be relevant once $\xi$ surpasses $\Lambda$.
It would seem homogeneous (thus SRC)
for the typical resolution of the system at criticality.
Surprisingly, this expected crossover from the LRC regime
to the SRC regime, where Harris criterion should apply,
has never been observed \cite{yoon,chan,larson}.
Experimental works \cite{larson} reveal that these changes
of criticality also occur in aerogels with relatively low porosities
($\varphi\!<\!90\%$) which do not seem to be fractal \cite{hasmy}.
These results appear to contradict WH criterion,
unless there exists LRC within the whole structure which is
not observable with standard techniques.
Furthermore, systematic violations of Josephson's hyperscaling
relation \cite{josephson,mefisher}
\be
\label{joseph}
d\nu=2-\alpha
\ee
have been evidenced in experiments involving estimations of $\alpha$ and $\nu$
within the same aerogel sample \cite{yoon,larson}.
These intriguing results have caused controversy.

In this letter, we report Monte Carlo simulations
of the 3DXY model confined in the pores of computer modeled aerogels.
These media are simulated through the on-lattice
diffusion limited cluster-cluster aggregation (DLCA) algorithm
\cite{meakin},
which has proved to reproduce well the structural features of aerogels
\cite{hasmy}.
We find a change of critical exponents $\nu$ and $\alpha$,
even at a relatively large concentration, consistent with experiments on
$^4\!$He confined in aerogels \cite{yoon,chan,larson}.
We argue that {\it hidden} LRC within DLCA aerogels
would explain these changes in criticality, so results do not really contradict
the WH criterion. Scaling consistency in experiments and simulations with
respect to theoretical predictions is established.

Simulated aerogels are built on simple cubic lattices of different sizes
($L\!=\!16-64$) using a standard DLCA model
with periodical boundary conditions (PBC) \cite{hasmy,meakin}.
A fraction $c$ of the total number $L^3$ of sites in the lattice
are initially occupied with $cL^3$ randomly disposed particles.
These are then allowed to, respectively, undergo Brownian motions,
until they irreversibly stick together when they enter in contact.
Aggregates are also allowed to perform Brownian motion, and
the procedure is repeated until all monomers and aggregates form
a single object.
This impurity distribution remains quenched in the box
in all that follows.

Empty sites of the lattice (aerogel pores) are filled up with planar spins
$\vec{\phi_i}\!=\!(\cos\theta_i, \sin\theta_i)$.
No interactions between spins and aerogel sites are considered.
Ferromagnetic interactions between nearest-neighbor pairs
$\langle i, j\rangle$ are described by the $XY$ Hamiltonian with PBC:
\be
-KH&=&\sum_{\langle i, j\rangle}K_{ij}\vec{\phi_i}\cdot\vec{\phi_j}\;,
\label{action}
\ee
\noindent
where $K\!=\!1/T$ is the coupling (inverse temperature, with $k_B\!\equiv\!1$);
$K_{ij}\!=\!0$ when the pair $\langle i, j\rangle$ includes at least
one aerogel site; and $K_{ij}\!=\!K$ elsewhere.
The magnetization squared (order parameter) is given by
$M^2=\|\langle\vec{\phi}\,\rangle\|^2= \big\|\sum_{i,j=1}^N
\vec{\phi_i}, \big\|^2/N^2 $,
where $N\!=\!(1\!-\!c)L^3$ is the effective number of $XY$-occupied sites.
Helicity modulus $[\langle\Upsilon_{\hat{\mu}}\rangle]_{L,c}$
has been determined using the Kubo formula \cite{moon,liteitel}
and averaged over the three directions
$\hat{\bf\mu}\!=\!\hat{\mbox{\bf x}}, \hat{\mbox{\bf y}}, \hat{\mbox{\bf z}}$
to improve statistics, assuming
that the model and disorder distributions are isotropic;
square brackets express the average over disorder realizations
for fixed $L$ and $c$, as will be stated below \cite{bernardet}.
Suitable statistics over disorder were considered: from
about $256$ realizations for $L\!=\!16$ to $16$ for $L\!=\!64$.
A cluster update has been used to minimize the effects of critical slowing down
\cite{machta},
defining one sweep as eight consecutive cluster flips.
Thermalization is reached after $2\!\times\!10^4$ sweeps
while $2\!\times\!10^6$
production sweeps are taken for further statistical analyses.
Reweighting techniques \cite{newman} and finite size
scaling (FSS) \cite{fisher} have been performed to obtain critical exponents.
Disorder averaging has been performed following Ref.\cite{bernardet}.
Critical couplings $K_c$ have been determined by
the crossing of Binder's magnetization fourth cumulant,
$[U_B]_{L,c}\!=\!1\!-\![\langle M^4\rangle]_{L,c}/3[\langle M^2\rangle]_{L,c}^2\,,$
which is size independent at $T_c$ for continuous phase transitions
\cite{binder,note1}.

Normalized critical temperatures are shown with circles in
Fig.1 as a function of impurity concentration $c$.
The dashed curve corresponds to the mean field approximation (MFAE),
$T_c(c)/T_c(0)=\langle z(c)\rangle/6$,
calculated through the average coordination number $\langle z(c)\rangle$
of the $XY$ sites.
This expression approaches our results well
at weak values of $c$.
Evident deviation at $c\!=\!0.2$ reflects that higher order terms,
beyond the mean field approach, should be taken in this case.
The random dilution expression (MFRD), $T_c(c)/T_c(0)\!=\!1\!-\!c$,
is illustrated with the dotted line for comparison.
The higher coordination number within aerogel pores
provide a rough geometrical explanation for
the curve concavity \cite{uzelac}.

Cross averages of $M^n$ with energy ($n\!\!=\!\!1,2,4$) are taken
to calculate logarithmic derivatives with
respect to the coupling $|\partial \ln(M^n)/\partial K|$ \cite{newman}. The
correlation length exponents $\nu$ have been determined using FSS,
$|\partial \ln(M^n)/\partial K|_{max}\!\sim\! L^{1/\nu}$.
Power law fits for $n \!=\! 2$ are depicted in Fig.2 (a)
for the pure system and for $c\!=\!0.05, 0.1, 0.2$.
Similar fits have been made for $n\!=\!1, 4\,$ in all cases
obtaining self-consistent results.
In agreement with experiments \cite{yoon,chan,larson},
exponents in confinement $[\nu\!\simeq\! 0.76(3)]$ differ satisfactorily
from the value obtained here for the pure system $[\nu\!=\!0.67(1)]$
which is consistent with previous works \cite{lipa}.
Fits have been performed including data for $L\!\geq\!L_{min}$, varying
$L_{min}$ from $16$ to $32$ observing no change in exponents (within error bars).
This reveals that there is no finite size crossover in the length scale studied,
and suggests that exponents of the 3DXY model confined into DLCA aggregates
effectively flow to a new universality class.
Exponent $\nu$ decreases slowly in the studied region $c\!=\!0.05\!-\!0.2$, though
detailed studies in denser aggregates should be made to confirm this tendency.

On the other hand, inflexion points from independent
$\Upsilon_L(K)$ extrapolated curves have been taken to determine averages
at pseudocritical couplings.
Plots for helicity modulus in Fig.2 (b)
($[\langle\Upsilon_{\hat{\mu}}\rangle]_{L,c}\!\sim\!L^{-\upsilon/\nu}$) show
that the finite size scaling exponent $\upsilon/\nu\!\simeq\!1$
is not affected by confinement.
Since helicity modulus is proportional to the superfluid density  
($\zeta\!=\!\upsilon$)
\cite{mefisher}, the hyperscaling relation (\ref{zeta}) is satisfied
for $d\!=\!3$ for all volume fractions tested.
Results suggest that the effective dimension of
aerogel pores occupied by $XY$ spins is Euclidean in fact.

Specific heat $C_L$ has been computed from energy fluctuations.
According to the scaling ansatz \cite{fisher}, it behaves as
$C_L^{max}\!-\!C_\infty^{max}\!\sim\!L^{\alpha/\nu}$.
The method developed by Schultka and Manousakis \cite{manousakis}
has been implemented to determine FSS exponents $\alpha/\nu$ directly.
While we obtain $\alpha/\nu\!=\!-0.019(2)$ $[\alpha\!=\!-0.013(2)]$
for the pure system,
an approximately constant value $\alpha/\nu\!\simeq\!-0.67$
yields $\alpha\!\simeq\!-0.5$ for the confined system,
in agreement with experiments on denser aerogels.
Writing down the relation (\ref{joseph}) as $\nu^*\!=\!2/(d\!+\!\alpha/\nu)$
we can compare both estimates of the correlation length exponent.
As shown in Fig.3, results with $\nu$ and $\nu^*$ differ in all $c\!\neq\!0$ cases.
This qualitatively agrees with experimental data
\cite{yoon,chan,larson} depicted in the figure for comparison.
The violation of hyperscaling relation (\ref{joseph}) 
observed in experiments \cite{yoon}
is reproduced here in terms of $\nu$ and $\nu^*$.
However, these results may require a more detailed analysis of 
finite-size effects on the
$\alpha/\nu$ exponent as in Ref.\cite{manousakis}.

Despite the agreement between our results and experimental data,
an apparent contradiction with the WH criterion remains to be explained.
As stated above, 
only LRC within disorder distribution could account for changes of exponents
observed in simulations and experiments \cite{weinrib,prudnikov}.
After detailed analyses on aerogels and DLCA aggregates,
only a finite region of power-law scaling is found for these objects \cite{hasmy}.
Nevertheless, previous studies show the existence of such LRC structures
within two-dimensional DLCA aggregates \cite{remi95}.
We carry out here a detailed study, in order to evidence
LRC structures within DLCA aggregates in $d\!=\!3$,
showing that non-fractality of aerogels at long scales
is only apparent.

First, we briefly define  the different objects studied
within the whole structure.
The {\it gelling cluster} (GC) is the first aggregate to reach
opposite sides of the box in any direction, $t_g$ is the time step at which
this occurs; {\it islands} are then defined as the smaller aggregates
not yet attached to the GC at $t_g$.
The {\it whole} DLCA is finally completed just a few steps
after $t_g$ when islands reach the GC.
Additionally, corresponding {\it backbones} \cite{herrmann} can be defined
for the GC (BBGC) and for the whole DLCA (BBDLCA)
as performed elsewhere for the two-dimensional case \cite{remi95}.
Using a box counting algorithm \cite{feder}, we obtain the average density
$n(r)\!=\!m(r)/r^d$ for $r\!\leq\!L/2$.
It is related to the two-point correlation function $g(r)\!\sim\!r^{-a}$
as $n(r)\!\sim\!r^{-(d-d_f)}$,  $d_f$ being
the fractal dimension of the object and $a\!=\!2(d-d_f)$.
Plots of $n(r)$ versus $r$ are sketched in Fig.4 for all structures described above
at volume fractions $c\!=\!0.05$ and $0.2$.

As observed in Fig.4,
LRC of gelling clusters (squares) appear hidden when one considers
the entire distribution of impurities at $t_g$,
{\it i.e.} GC+islands (diamonds) whose plots
coincide with those of whole DLCA (triangles).
The GC alone and respective BBGC (circles) scale at distances comparable to
the lattice size ($L/2$ at PBC).
The average inverse slope of plots from Fig.2 (a)
has been depicted for a qualitative comparison (dashed line).
Arrows indicate the crossover to homogeneous regimes, evident for
the whole DLCA and the GC+islands distributions,
while fractal regimes of GC and BBGC span all scales.
In conclusion from these results, the short-ranged fractal regime of DLCA 
would prevent the structure as a whole 
to be relevant for criticality of the 3DXY model.

Authors in Refs.\cite{yoon,chan} noticed the same paradox
for the $\lambda$ transition of $^4\!$He in aerogels:
standard measures evidence a finite cutoff
for the fractal regime \cite{hasmy} beyond which
aerogels appear homogeneous.
If disorder were strictly homogeneous at large scales, then
the tail of its correlation function would decay exponentially whence
Harris criterion applies: a crossover to the SRC regime would be
expected as $T\rightarrow T_c$.
However, such a crossover is never observed:
exponents rest still different from those of pure $^4\!$He 
as $T_c$ is closely approached in experiments,
and this is due to LRC according to the WH criterion.
Note that islands have a homogeneous distribution
at $t_g$, so when they finally reach the GC at random sites,
the whole DLCA appears homogeneous in spite of the underlying 
LRC fractal within.
If we take the 3DXY model in the presence of a
random distribution of impurities as the original critical system,
it has the same exponents as the pure system. Then, it
should certainly be affected by a LRC distribution of vacancies (GC)
which actually defines the observed critical behavior.
WH considered the correlation function as a sum of power-law terms,
$g(r)\!\sim\!\sum_i\! c_ir^{-a_i}$ \cite{weinrib},
to show that it is the dominant term of $g(r)$
(smallest $a_i$) which defines criticality.
This decomposition of the correlation function, together with
the analysis of structures sketched above, confirms that the change of
criticality observed in the 3DXY model is not in contradiction 
with the WH criterion.

To conclude, we have found clear changes of criticality in the 3DXY
model when confined in DLCA aerogels at different porosities.
Evidence of long-range correlated structures
within DLCA aggregates in three dimensions has been presented,
applicable to real silica aerogels as well.
This correlation functions analysis should clarify the reason these changes
of criticality occur in the 3DXY model in DLCA, and why the crossover to 
bulklike behavior of superfluid $^4\!$He, confined in aerogels, 
has not been observed.

We are grateful to the Centre Informatique National de l'Enseignement 
Sup\'erieur (CINES, France) and CeCalc-ULA (Venezuela) for computer facilities.
Discussions with E. Anglaret, E. Medina, B. Berche, C. Goze, T. Woignier, and
the support from the Venezuelan-French PCP exchange program are also kindly
acknowledged.

\newpage
\relax

\newpage
\begin{figure}[!th]
\begin{center}
\includegraphics[width=16cm]{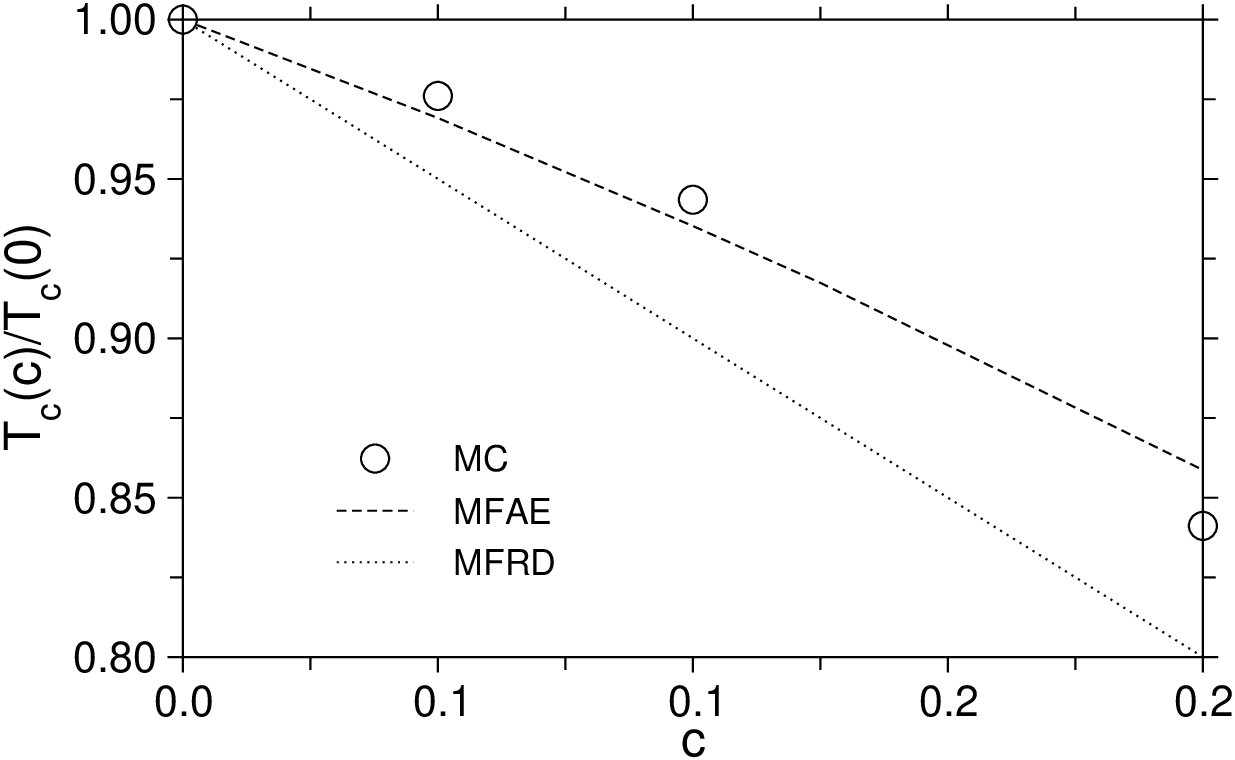}
\caption{\label{Tc}
Relative critical temperature $T_c(c)/T_c(0)$ (circles)
compared to the mean-field approximation (dashed line).
The dotted line corresponds to the random dilution case.
}
\end{center}
\end{figure}

\newpage
\begin{figure}[!th]
\includegraphics[width=16cm]{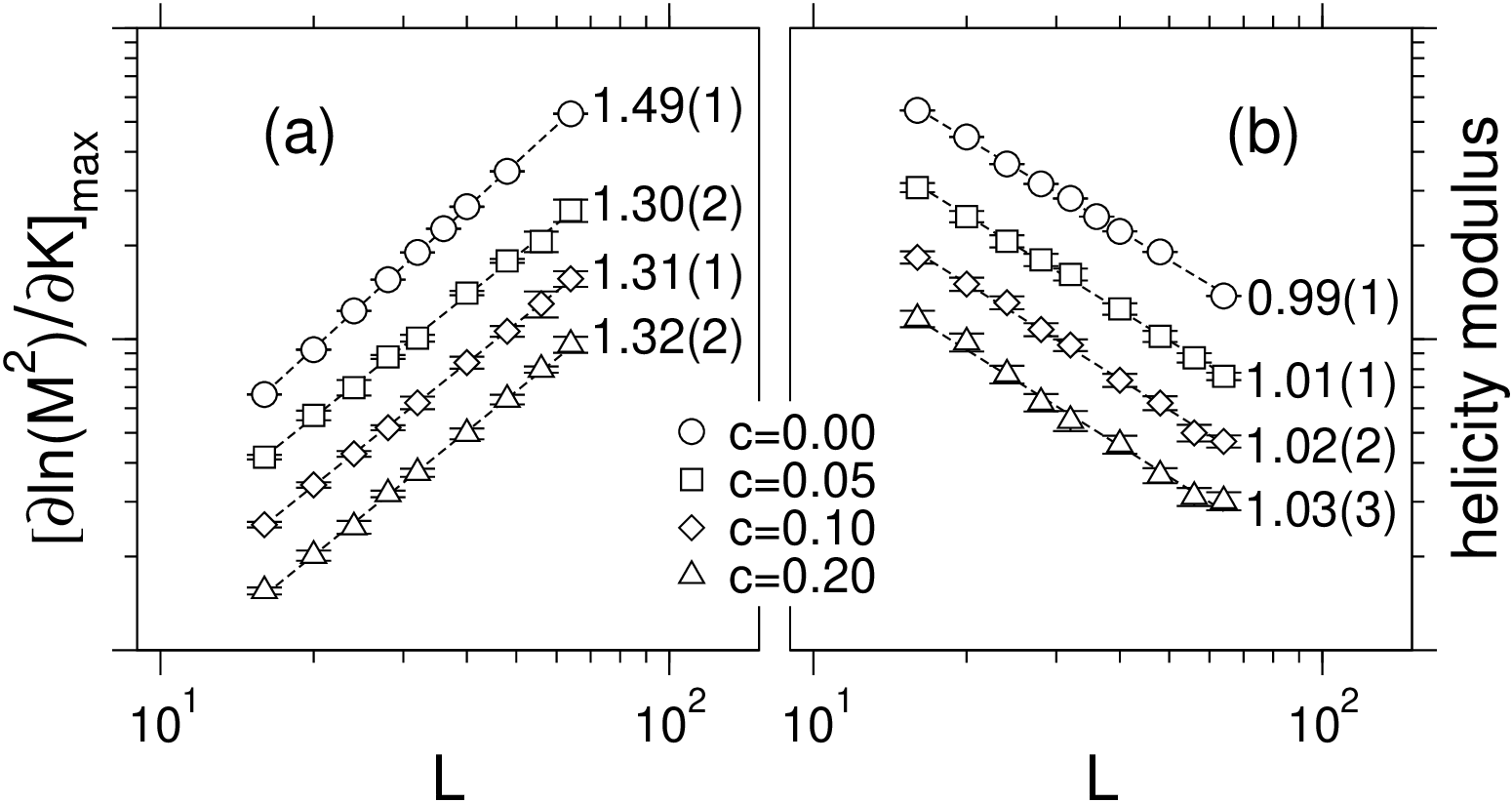}
\caption{
Finite size scaling of maximum logarithmic derivatives (a)
and helicity modulus (b), for the pure 3DXY and
$c\!=\!0.05,\,0.1,\,0.2 $ as indicated.
Plots have been displaced vertically for clarity
(vertical axes are in arbitray units).
Corresponding FSS slopes are indicated on each graph.}
\end{figure}
\newpage
\begin{figure}[!th]
\includegraphics[width=16cm]{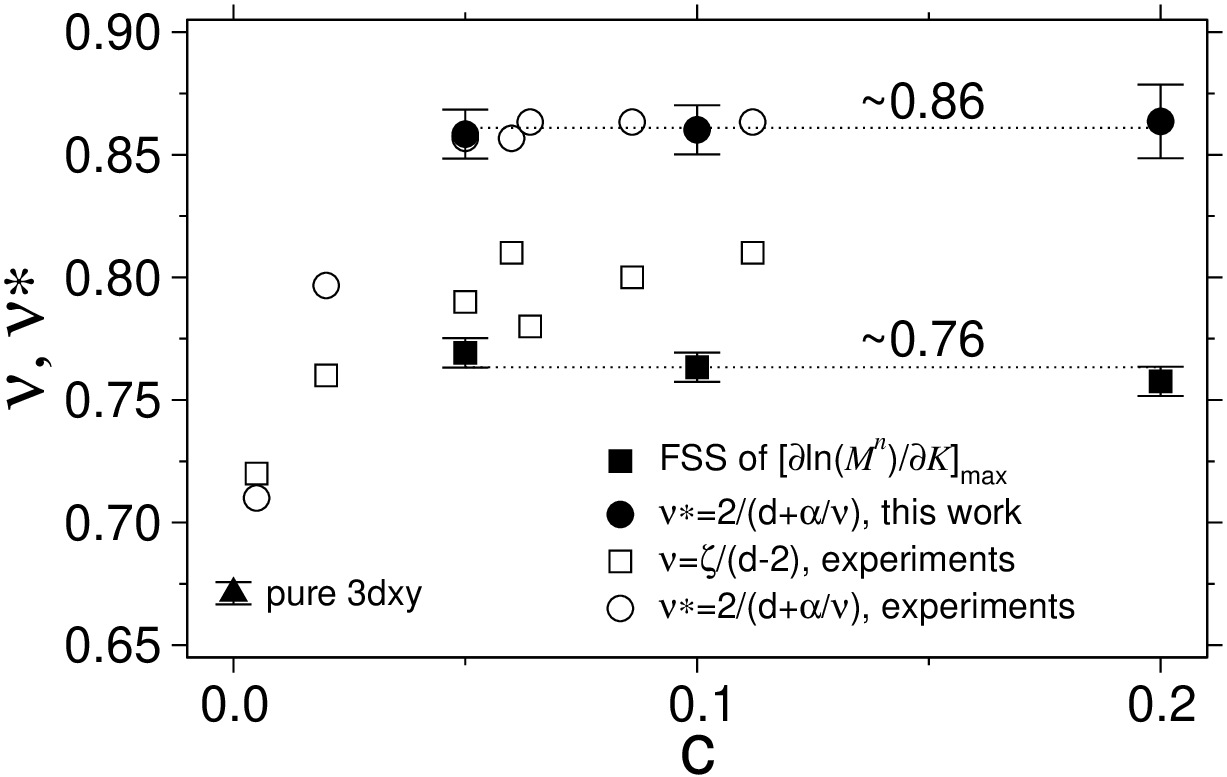}
\caption{
Exponents $\nu$ and $\nu^*$ {\it vs.} aerogel volume fraction $c$,
taken from FSS fits in Fig.2 (black squares), and from FSS of
maximum specific heat (black circles), using Josephson's relation
$\nu^*\!=\!2/(d\!+\!\alpha/\nu)$.
Results from Refs.\cite{yoon,chan,larson},
depicted here for comparison (corresponding open symbols).
Dotted lines show average values for $c\!\neq\!0$ as indicated.
}
\end{figure}
\newpage
\begin{figure}[!th]
\includegraphics[width=16cm]{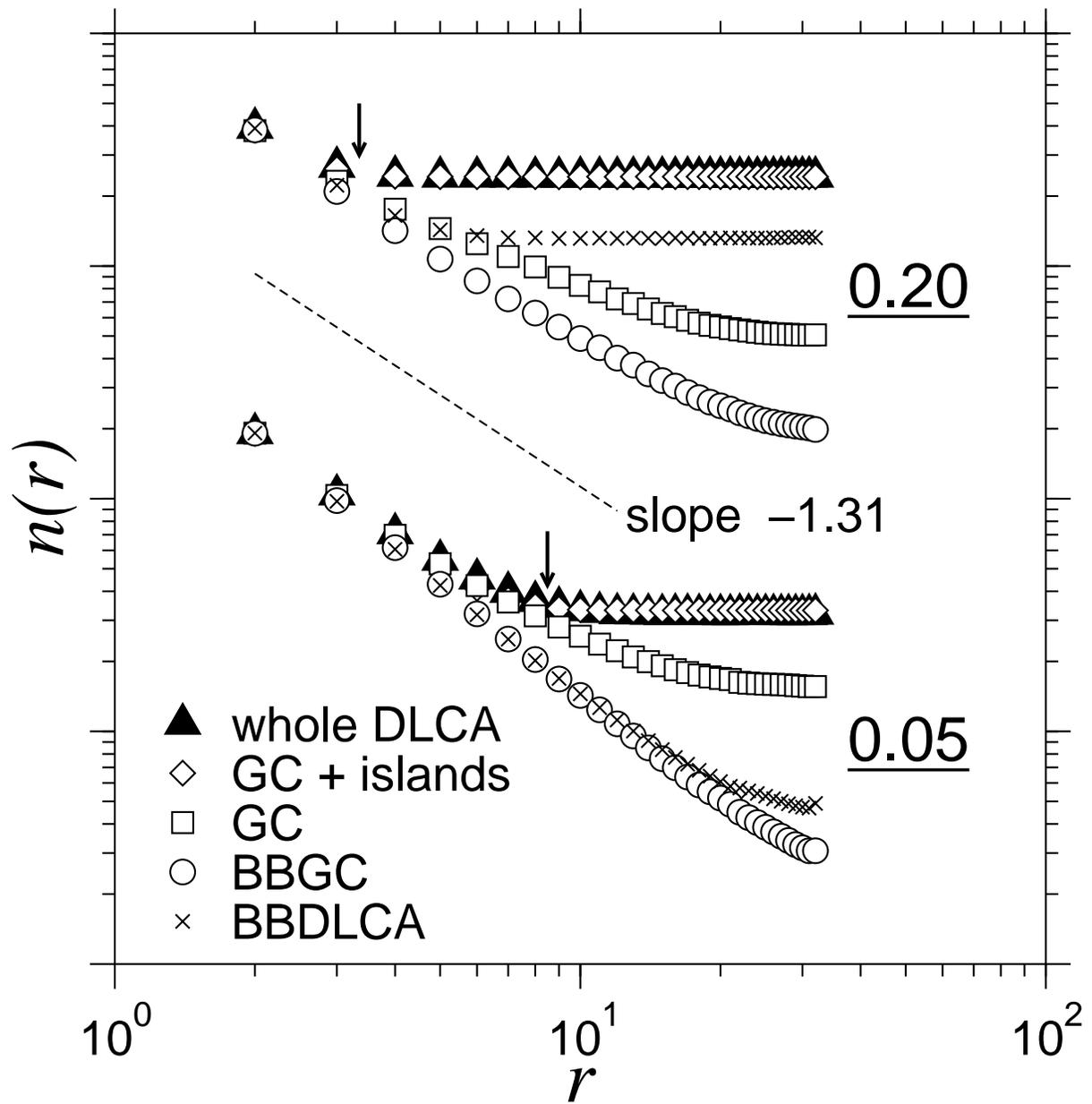}
\caption{
Plot of $n(r)$ versus $r$ for structures within DLCA,
as indicated, at $c=0.05$ and $0.2$.
Arrows indicate crossover to homogeneous regimes in DLCA.
Lattice sizes are $L=64$, plots result from averaging over 32 realizations.
The dashed line corresponds to the slope $-1/\nu$.}
\end{figure}


\begin{thebibliography}
\bibitem{note1}
Critical couplings $K_c$ have also been estimated through FSS
of the positions of maximum logarithmic derivatives,
$K^*_c(L) \simeq K_c + aL^{-1/\nu}$, giving self consistent results.
\end{thebibliography}
\end{document}